\documentclass[manuscript]{acmart}
\usepackage{caption}
\usepackage{subcaption}
\usepackage{pifont}% http://ctan.org/pkg/pifont
\usepackage{xac}

\AtBeginDocument{%
  \providecommand\BibTeX{{%
    \normalfont B\kern-0.5em{\scshape i\kern-0.25em b}\kern-0.8em\TeX}}}

\begin{document}

\title{OralViewer: 3D Demonstration of Dental Surgeries for Patient Education with Oral Cavity Reconstruction from a 2D Panoramic X-ray}

\author{Yuan Liang}
\affiliation{%
  \institution{University of California, Los Angeles}
%   \streetaddress{Anonymous for Review}
%   \city{Anonymous for Review}
  }
\email{liangyuandg@g.ucla.edu}

\author{Liang Qiu}
\affiliation{%
  \institution{University of California, Los Angeles}
%   \streetaddress{Anonymous for Review}
%   \city{Anonymous for Review}
  }
\email{liangqiu@g.ucla.edu}

\author{Tiancheng Lu}
\affiliation{%
  \institution{University of Pittsburgh}
%   \streetaddress{Anonymous for Review}
%   \city{Anonymous for Review}
  }
\email{til62@pitt.edu}

\author{Zhujun Fang}
\affiliation{%
  \institution{University of California, Los Angeles}
%   \streetaddress{Anonymous for Review}
%   \city{Anonymous for Review}
  }
\email{floraqa@g.ucla.edu}

\author{Dezhan Tu}
\affiliation{%
  \institution{University of California, Los Angeles}
%   \streetaddress{Anonymous for Review}
%   \city{Anonymous for Review}
  }
\email{dezhantu@gmail.com}

\author{Jiawei Yang}
\affiliation{%
  \institution{University of California, Los Angeles}
%   \streetaddress{Anonymous for Review}
%   \city{Anonymous for Review}
  }
\email{jiawei118@g.ucla.edu}

\author{Tiandong Zhao}
\affiliation{%
  \institution{University of California, Los Angeles}
%   \streetaddress{Anonymous for Review}
%   \city{Anonymous for Review}
  }
\email{zhaotiandong@ucla.edu}

\author{Yiting Shao}
\affiliation{%
  \institution{University of California, Los Angeles}
%   \streetaddress{Anonymous for Review}
%   \city{Anonymous for Review}
  }
\email{shaobruin0011@ucla.edu}

\author{Kun Wang}
\affiliation{%
  \institution{University of California, Los Angeles}
%   \streetaddress{Anonymous for Review}
%   \city{Anonymous for Review}
  }
\email{wangk@ucla.edu}

\author{Xiang `Anthony' Chen}
\affiliation{%
  \institution{University of California, Los Angeles}
%   \streetaddress{Anonymous for Review}
%   \city{Anonymous for Review}
  }
\email{xac@ucla.edu}

\author{Lei He}
\affiliation{%
  \institution{University of California, Los Angeles}
%   \streetaddress{Anonymous for Review}
%   \city{Anonymous for Review}
  }
\email{lhe@ee.ucla.edu}

\renewcommand{\shortauthors}{Yuan Liang, et al.}


% \xac{in the title, should we highlight that the reconstruction is from a 2D x-ray?}
\begin{abstract}
Patient's understanding on forthcoming dental surgeries is required by patient-centered care and helps reduce fear and anxiety. 
Due to the gap of expertise between patients and dentists, conventional techniques of patient education are usually not effective for explaining surgical steps. 
% 3D visualization and simulation of surgeries offer a promising possibility to promote patient's understanding, however, to the best of our knowledge, no solution is available for dental clinics. 
In this paper, we present \textit{OralViewer}---the first interactive application that enables dentist's demonstration of dental surgeries in 3D to promote patients' understanding. 
\textit{OralViewer} takes a single 2D panoramic dental X-ray to reconstruct patient-specific 3D teeth structures, which are then assembled with registered gum and jaw bone models for complete oral cavity modeling. 
During the demonstration, \textit{OralViewer} enables dentists to show surgery steps with virtual dental instruments that can animate effects on a 3D model in real-time. 
A technical evaluation shows our deep learning based model achieves a mean Intersection over Union (IoU) of 0.771 for 3D teeth reconstruction. 
A patient study with 12 participants shows \textit{OralViewer} can improve patients' understanding of surgeries. 
An expert study with 3 board-certified dentists further verifies the clinical validity of our system.

\end{abstract}




\begin{teaserfigure}
  \includegraphics[width=\textwidth]{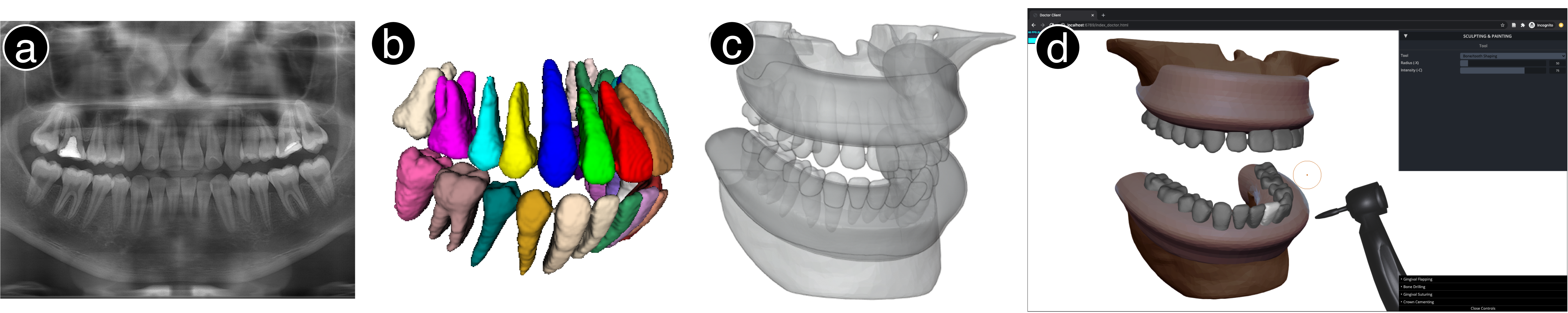}
  \caption{Overview of \textit{OralViewer}. The system takes a patient's 2D X-ray as input (a), and reconstructs the 3D teeth structure (b) with a novel deep learning model. The system then generates the complete oral cavity model (c) by registering the pre-defined models of jaw bone and gum to the dental arch curve. Finally, a dentist can demonstrate the forthcoming surgeries to a patient by animating the steps with our virtual dental instruments (d). } 
  \label{fig_teaser}
\end{teaserfigure}

\maketitle

\section{Introduction}
% patient communication 
Patient-dentist communication is a core requirement of patient-centered care \cite{asimakopoulou2014patient}. 
According to \cite{travaline2005patient,mills2014patient,stewart1995effective,rozier2011dentist}, patients who understand their dentists and procedures are more likely to follow medication schedules, feel satisfied about treatments and have better oral health outcomes. 
% fear  
Moreover, many patients about to undergo oral surgeries can experience anxiety --- up to every fourth adult reported dental fear \cite{oosterink2009prevalence} and it has been cited as the fifth-most common cause of anxiety \cite{agras1969epidemiology} among all kinds of anxiety. 
% \xac{among all kinds of anxiety? or medically related anxiety?} 
To manage the dental fear, one solution is to unveil the surgical steps with patient education to decrease patients' fear of the unknown \cite{johnson1984stress,armfield2013management,bailey2010strategies}. 
Previous studies have observed shorter duration of surgery \cite{elter1997assessing}, lower level of post-operative pain \cite{kain2000preoperative}, and smoother recovery \cite{kazancioglu2015does} with a reduced dental fear. 

% current oral 
Currently, dentists mostly perform pre-operative patient education via verbal explanation, and possibly with the aid of hand-drawn diagrams \cite{burghardt2018non}, audiovisual slides \cite{choi2015effect} and video clips \cite{kazancioglu2015does}. 
% 3d important 
Meanwhile, the recent advent of 3D demonstration, which illustrates complex procedures with dynamic visuals in 3D, has shown an increasing potential in patient education since it is more intuitive and complete than verbal description and static images.
% \xac{maybe a figure that show different existing methods of demosntrations?} 
% non-oral 
Indeed, existing studies have explored the 3D demonstration for cardiac surgeries \cite{capuano2019computational,loke2019abnormal}, condylar reconstruction \cite{yang2011computer} and pancreatectomy \cite{endo2014patient} to assist the pre-operative communications.  
% however no 
However, to the best of our knowledge, the use of 3D demonstration for dental clinics is still an underexplored area.
% \xac{define what is `3d demonstration' and why do we need it given there is 3d simulation?s}

To fill this gap, we present \textit{OralViewer}, a web-based system to enable dentists to virtually demonstrate dental surgeries on 3D oral cavity models for patient education. 
To inform the design of \textit{OralViewer}, we interviewed three dentists and elicited key system requirements from a clinical point of view: (\textit{i}) providing a patient-specific 3D teeth model, 
% \xac{to dentists, does it matter where the 3d model come from?}
(\textit{ii}) modeling the complete oral cavity of teeth, gums and jaw bones,
% \xac{do you mean modeling everything in the oral cavity?}
% \xac{what is `comprehensive oral cavity'? you mean with a complete view of the oral cavity?}
, and (\textit{iii}) demonstrating surgery steps using simple operations. 
In terms of 3D oral cavity modeling, \textit{OralViewer} goes beyond existing work \cite{badash2016innovations,lin2019introduce,welker2014urogynecology} that extracts a patient's anatomy models from high-cost 3D scanning, \textit{e.g.}, computerized tomography (CT) scans.
Instead, we enable the generation of 3D models from a single 2D panoramic X-ray image with a novel deep learning model. 
This approach lowers the barrier of obtaining a 3D model as the 2D panoramic X-ray is the most common modality in dentistry and the only required imaging for many dental surgeries \cite{langland2002principles}.
In terms of surgery demonstration, \textit{OralViewer} implements virtual dental instruments that are simple to operate with a mouse and illustrative with real-time effects on oral cavity models for patients to understand. 
Figure \ref{fig_teaser} shows the workflow:
the system first takes a patient's panoramic X-ray to generate the 3D teeth structure (a$\rightarrow$b); 
then pre-defined gum and jaw bone models are registered to the dental arch, and assembled with the teeth structure for the complete oral cavity model (c);   
finally, a dentist operates virtual dental instruments on the reconstructed oral cavity to demonstrate a forthcoming surgery to a patient (d). 

We validate \textit{OralViewer} for the demonstration of two common dental surgeries: crown lengthening and apicoectomy. 
Note that the design and implementation of \textit{OralViewer} (\eg the 3D reconstruction process and virtual operation techniques) are expected to generalize to other dental surgeries as well. 
These two surgeries were selected because each of them involves multiple
% \xac{what do you mean by non-trivial?} 
steps and and requires various commonly-used dental instruments, thus are ideal for testing the demonstration capability of \textit{OralViewer}. 
We conducted three evaluations: 
(\textit{i}) A technical evaluation of 3D teeth reconstruction from 2D panoramic X-ray shows our model achieves an average IoU of 0.771$\pm$0.062. 
(\textit{ii}) A study with 12 patient indicates that our system leads to patients' improved understanding of dental surgeries. 
(\textit{iii}) An expert study with 3 board-certificated dentists suggests that the demonstration using our system is clinically valid, can improve the efficiency of patient education, yet there remain areas for improvement in the ease of operation of the virtual tools.
% interaction modification can be made to further improve operations.  

\textbf{Contributions} of this paper include:
\vspace{-1em}
\begin{itemize}
    \item \textit{OralViewer}~---~the first solution that enables 3D demonstration of dental surgeries for patient education; 
    \item 3D modeling~---~the first 3D reconstruction technique of oral cavity from a single 2D panoramic X-ray; 
    \item Evaluation~---~a patient study and an expert study validate the feasibility and usability of educating patients with 3D simulative demonstration. 
\end{itemize}

\section{Background}
In this section, we briefly introduce the common steps in typical apicoectomy and crown lengthening surgeries. 
Detailed information about both dental surgeries can be found in \cite{gutmann2010problem}, and more descriptions with figures are included in Supplementary Material 1.1.
% \ref{supp_background}.

\textbf{Apicoectomy} is the removal of the root tip and surrounding tissues of a tooth with periapical inflammation.
A dentist first performs a periodontal flap --- incises and flaps the gum tissue for unveiling the underlying bone structure with scalpels.  
% The flap design considers multiple factors, \eg the position of the tooth and extent of lesion. 
Next, the apex is exposed by creating an peripheral opening on the buccal (jaw) bone with a round bur and a steady stream of saline solution. 
After that, the inflamed root tip can be resected with a handpiece, followed by filling material into the tooth cavity created to seal it. 
Then, bone grafting materials can be injected into the jaw bone hole for rehabilitation, and finally the periodontal flap being repositioned and sutured.

\textbf{Crown Lengthening} can be applied for restoring cavities and tooth fractures that happen below the gum tissue.
To start, a dentist incises and flaps the gum tissue to unveil the target structure.
Next, the jaw bone's height at a surrounding area is often reduced with a bur, in order to support the repositioned gum in a lower position below the cavity/fracture.
Then, the cavity is removed (or fracture shaped) with a handpiece, and restored with grafting materials. 
For better protecting the tooth, the restored crown is sometimes further shaped (with handpieces), and cemented with an artifact crown.

\section{Related Work} 
In this section, we first review existing work on 3D surgical visualization and simulation.
Then, we summarize the 3D-based Computer-Aided Design (CAD) technologies for dentistry. 
We also include a review on the deep learning based algorithms for 3D reconstruction from a single 2D view. 

\subsection{Surgical Visualization and Simulation}
Conventional techniques of delivering education to patients through verbal instructions may not be effective to explain surgical procedures due to the educational barriers between the patients and the clinicians. 
Researches have shown 3D anatomy visualization can improve the patients’ understanding of surgeries, where example systems include those for abdominal \cite{lin2019introduce}, cardiac system \cite{whitman2017visualization,olivieri2018novel}, and more \cite{welker2014urogynecology}. 
% Researches have shown that multimedia resources can improve the patients’ understanding of surgeries that they are about to undergo \cite{jones2001more,mulsow2012beyond}.
% % visulzation 
% For example, systems that 3D visualize the anatomies of interests, \textit{e.g.}, abdominal \cite{lin2019introduce}, cardiac system \cite{whitman2017visualization,olivieri2018novel}, and pelvic organ prolapse \cite{welker2014urogynecology}, have been developed to assist explaining the conditions and procedures. 
% simulaiton
Moreover, with the advent of computer graphics, interactive manipulation of virtual 3D models has shown to help patients acquire a more satisfactory level of knowledge \cite{prinz2005advantage,endo2014patient,eschweiler2016biomechanical}. 
% For example, there have been simulations of ophthalmic surgery \cite{prinz2005advantage}, abdominal surgery \cite{endo2014patient}, and wrist joint therapy \cite{eschweiler2016biomechanical}. 
% vr  
Recently,
% systems \footnote{\url{https://vsi.health/en/holomedicine/vsi-patient-education/}} and 
studies 
% \cite{shirk2019effect,satava1993virtual,delp1990interactive}
\cite{shirk2019effect,satava1993virtual}
have also incorporated virtual reality (VR) to enable more intuitive anatomy viewing. 

% ct -> 2d 
In comparison, \textit{OralViewer} is different from all the aforementioned work on two aspects. 
First, all the 3D anatomy models used are either captured from 3D scanning, \textit{e.g.}, CT, or utilizing a one-size-fits-all standard model.  
However, considering the limited availability of 3D imaging for dental surgeries, \textit{OralViewer} generates the detailed oral cavity model from a single 2D panoramic X-ray with a novel 3D reconstruction algorithm.
Second, to the best of our knowledge, no existing study has enabled the 3D demonstration of dental surgeries for patient education, which we explore in the design and implementation of \textit{OralViewer}. 
% Thus, the simulation design of dental procedures is still open to explore.  

\subsection{Computer-Aided Design (CAD) for Dentistry}
% tools to design 
CAD tools have been widely applied in dentistry to improve the design of dental restorations, \textit{e.g.} crowns, dental implants and orthodontic appliances \cite{davidowitz2011use,rekow1987computer}. 
% examples 
Specifically, models of patients' oral cavity are created from digital 3D scanning, based on which dentists produce a virtual design of restorations for manufacturing \cite{oen2014cad}. 
% \xac{what is `restorative device' and `milling process'}
% Such CAD tools have proven to improve the restoration accuracy and shorten manufacturing periods \cite{oen2014cad}. 
% cannot illustation 
However, all the CAD tools are aimed to guide a clinician through restoration designing \cite{strub2006computer, taneva20153d}, rather than patient education, which is the focus of \textit{OralViewer}. 
% Thus, the aforementioned tools have few considered the oral cavity visualization and surgical procedure simulation for patient's understanding. 
Thus, oral cavity visualization and surgical step simulation have not been considered in existing CAD tools when it comes to patient education. 
% \xac{how about those videos that animate dental procedures?}
% ct -> 2d 
Moreover, 3D imaging of patient's oral cavity, \eg CT and intra-oral scanning \cite{davidowitz2011use,raja2016computer}, is almost always required by the existing CAD tools.
% for the dental restoration design.
In contrast, \textit{OralViewer} reconstructs the patient's 3D oral cavity from the 2D panoramic X-ray, which is one of the most common imaging modalities in dentistry \cite{whaites2013essentials}, in order to enable the application of the system for a wide range of dental surgeries.

\subsection{Single-View 3D Reconstruction}
% define 
Single-view 3D reconstruction aims at generating the 3D model of an object based on a single 2D projection of it.  
% deep learning 
Currently, deep Convolutional Neural Networks (ConvNets) based methods have achieved the highest accuracy in various benchmarks by using both low-level image cues, \textit{e.g.}, texture, and high-level semantic information \cite{tatarchenko2019single,xian2018monocular}. 
% categories 
According to the representation type of 3D outputs, most existing work can be categorized into: (\textit{i}) voxel-based \cite{choy20163d,girdhar2016learning,tatarchenko2017octree}, (\textit{ii}) mesh-based \cite{wang2018pixel2mesh,groueix2018papier,mescheder2019occupancy}, and (\textit{iii}) point-cloud-based \cite{tatarchenko2016multi,lin2017learning}.
A detailed review of the above categories of methods can be found in \cite{tatarchenko2019single}.
% Voxel-based work estimates a voxel occupancy grid for indicating if voxels are within the space of an object, mainly by first encoding the 2D view into a latent representation, and then restoring the 3D shapes with 3D up-convolutions \cite{choy20163d,girdhar2016learning,tatarchenko2017octree}.
% Mesh-based methods output the prediction of an object's surface by deforming spheres into a desired shape \cite{wang2018pixel2mesh,groueix2018papier,mescheder2019occupancy}. 
% Point-cloud-based methods predict a set of points representing the external surfaces of an object, typically either by direct regression \cite{tatarchenko2016multi,lin2017learning} or fusing the generated multi-view depth maps \cite{fan2017point}.
% uniqe challeges: 
Our work targets at generating the voxel-based representation of teeth volumes, which estimates a voxel occupancy grid for indicating if voxels are within the space of an object. 
The representation selection mainly considers the need for smooth and closed-surface models, even with the presence of complex typologies on occluded surfaces.
A few existing work \cite{abdelrahim2012realistic,abdelrehim20132d,liang2020x2teeth} explored teeth reconstruction from X-ray, however, they either targeted at single tooth or worked with synthesized images only, which cannot serve our propose of patient-specific modeling and demonstration. 
To the best of our knowledge, ours is the first work on exploring 3D reconstruction of teeth structures from clinical 2D panoramic X-rays. 

\section{Formative Study} 
To understand the system requirements of \textit{OralViewer} from a clinical point of view, we conducted interviews with three dentists (two female and one male). 
We started by asking for the method he/she applies to perform patient education.
% , and what types of information they need to communicate with the patients. 
We then described the motivation and goal of \textit{OralViewer}, emphasizing on using a 3D model to visualize and simulate surgical steps to laymen patients. 
We gathered and built on dentists’ feedback to formulate the below system requirements that ensure clinically-valid demonstration and user-friendliness. 

% \begin{itemize}
% \item 
\textbf{R1. Providing patient-specific teeth model.}
% patient specific
Surgical steps, \textit{e.g.} how a fractured tooth is extracted or repaired, often depend on individual's teeth condition.
Thus, patient-specific teeth model should be provided to make demonstrations contextualized to the patient's conditions. % without misleading patients by inconsistent conditions. 
% know 
Moreover, compared to panoramic oral X-ray, 3D screening of oral cavity is not a standard practice for the clinical diagnosis of many common surgeries, \textit{e.g.} apicoectomy, root canal treatment, and crown lengthening, for its higher radiation and cost. 
% ct is not 
As such, it is preferred to generate a patient's 3D teeth model from his/her 2D X-ray image to enable the widely available application of the system.

% \item 
\textbf{R2. Modeling complete oral cavity.}
Both the target oral structure of a surgery and its nearby anatomies need to be incorporated into a surgical demonstration.
For example, when dentist removes a root tip in apicoectomy, procedures on other structures should be simulated as well, \textit{e.g.} some gum tissue will be lifted from an area near the root tip and some surrounding bone will be removed. 
Thus, to help patients understand what to expect in a surgery, complete oral cavity including teeth, gum, and jaw bones should be modeled. 

% \item 
\textbf{R3. Demonstration in simple operations.}
% \xac{now you seem to be using demonstrate, animate and simulate interchangeable--should try to stick to one word}
Dentists consider it important to show for each surgery step: (\textit{i}) how the step is performed --- illustrating the applied instruments, and (\textit{ii}) what happens in the step --- animating the dental structure changes upon the application of instruments. 
Moreover, the demonstration should be carried out by dentists using simple interaction techniques, which is more important than having to achieve realistic effects with a high fidelity.
% For example, to demonstrate removing a piece of gum tissue, dentists prefer a simple operation (\eg drawing an inexact line on the gum's 3D model, which removes a predefined shape from it) rather than simulatively performing a fine incision as in an actual surgery.
For example, to demonstrate shaping a tooth with a dental handpiece, dentists prefer a simple operation, \eg pressing and dragging the cursor on desired places of a tooth with customizable effect to simulatively perform a grinding as in an actual surgery.
% \xac{reviewer might question that such oversimplification might mislead the patient---did the dentist concern about this?}
% \end{itemize}

\section{OralView Design and Implementation} 
Guided by the aforementioned requirements, we designed and implemented \textit{OralViewer} for 3D demonstration of dental surgery to patients.
The \textit{OralViewer} consists of two cascaded parts: \one a 3D reconstruction pipeline for generating a patient's oral cavity from a single 2D panoramic X-ray, and \two a demonstration tool for dentist's animating steps on the 3D model with virtual dental instruments. 
% \xac{need rephrased}

\subsection{3D Reconstruction of Oral Cavity} 
\textit{OralViewer} reconstructs a complete oral cavity model consisting of teeth, gum and jaw bones, all of which are vital for the demonstration of surgical procedures (\textbf{R2}). 
Importantly, to reflect the patient-specific dental condition, we estimate the patient's 3D teeth structures from a single 2D panoramic X-ray with an end-to-end trainable deep ConvNet model (\textbf{R1}). 
Since the 3D structures of soft tissues, \textit{i.e.}, gum, and jaw bones cannot be well reflected from X-ray \cite{haring2000dental}, their 3D templates are pre-defined and can be registered to tailor for specific patients' oral anatomy.  
% The template models are registered to a patient's dental arch curve to tailor for the oral condition, and assembled with the estimated teeth structures for the complete oral cavity model.

% \subsubsection{3D Reconstruction of Teeth}

\begin{figure*}
  \includegraphics[width=0.8\textwidth]{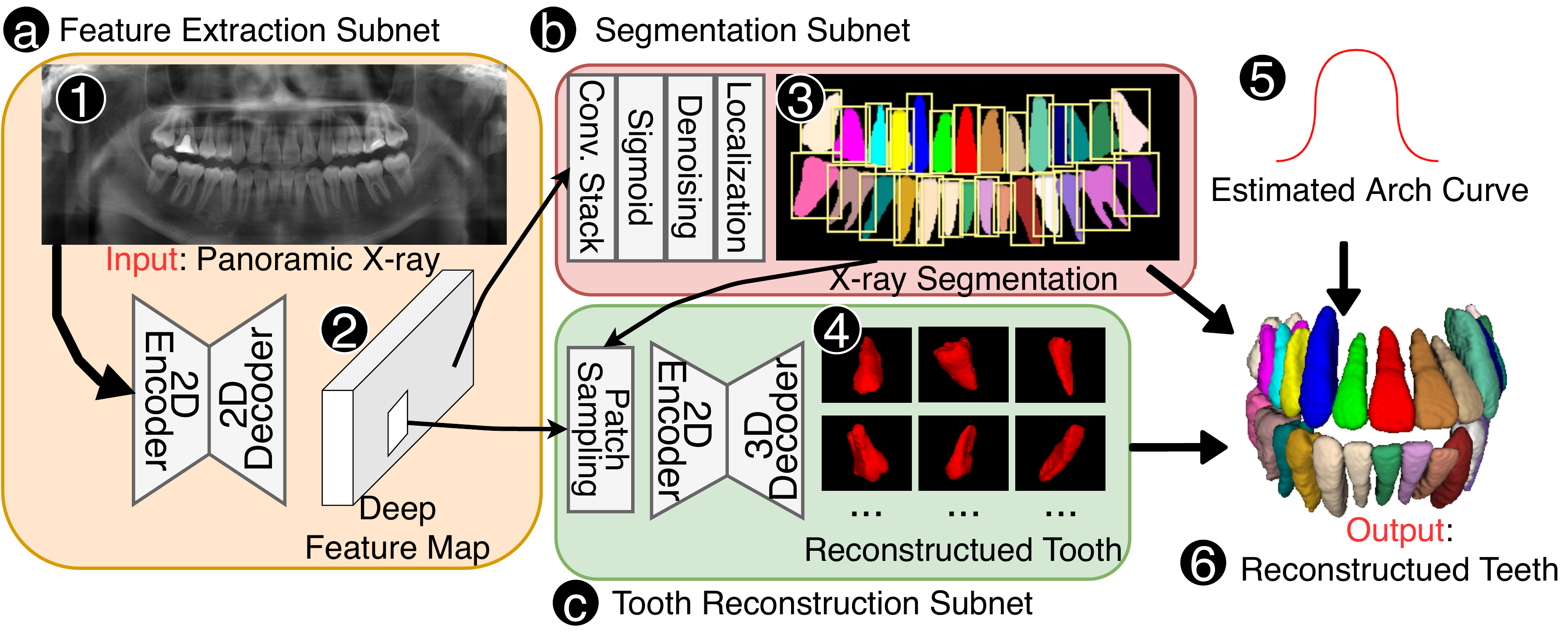}
  \caption{Overall architecture for reconstructing 3D tooth structures from a panoramic X-ray. 
  % \xac{where does the estimated arch curve come from?}
  } 
  \label{fig_model}
  \end{figure*}

\begin{figure}[t]
  \centering
  \begin{subfigure}{.35\textwidth}
    \includegraphics[width=0.9\linewidth]{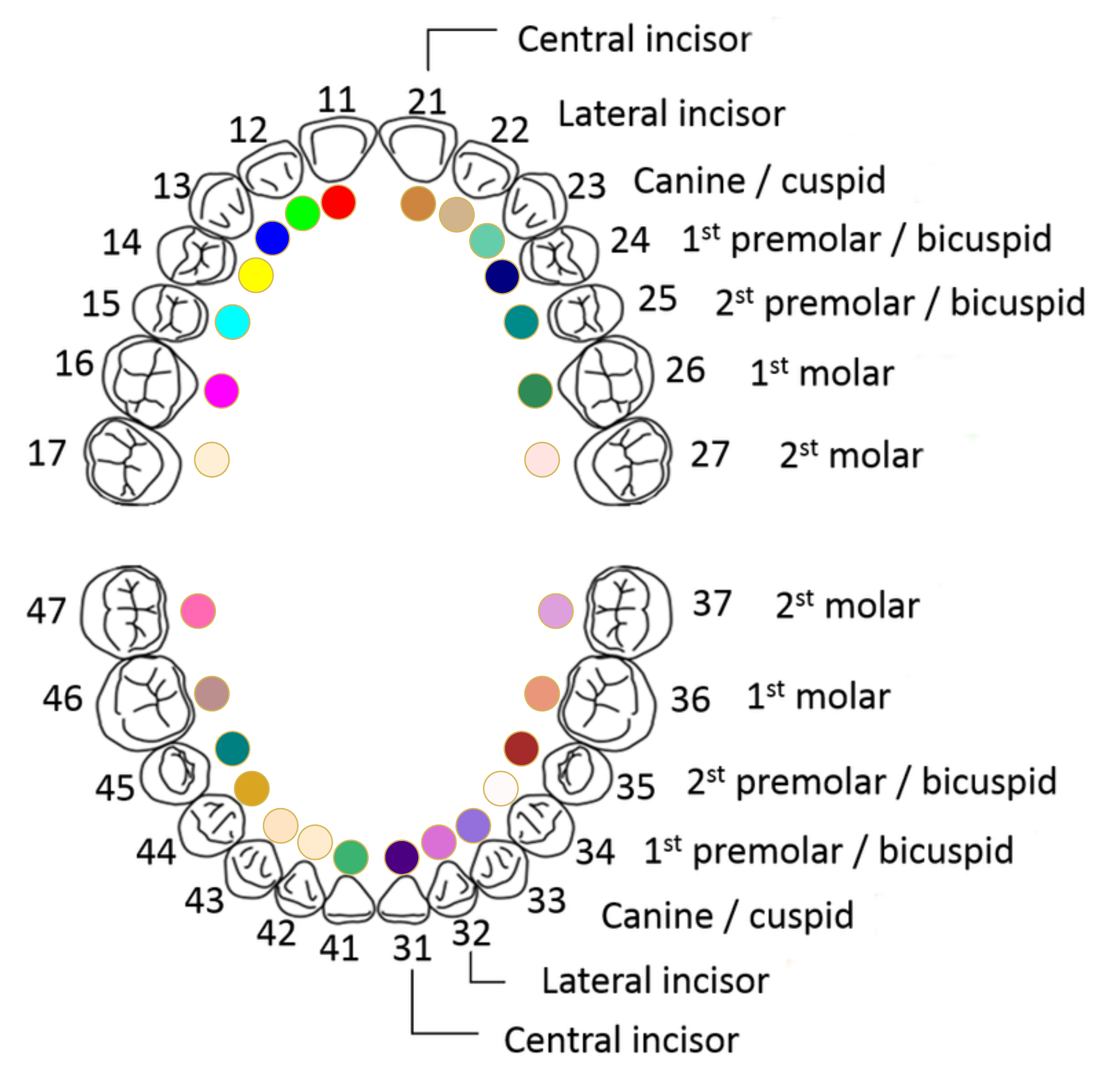}
    \caption{
    }\label{fig_numbering}
  \end{subfigure}%
  \hfill
  \begin{subfigure}{.65\textwidth}
    \includegraphics[width=0.9\linewidth]{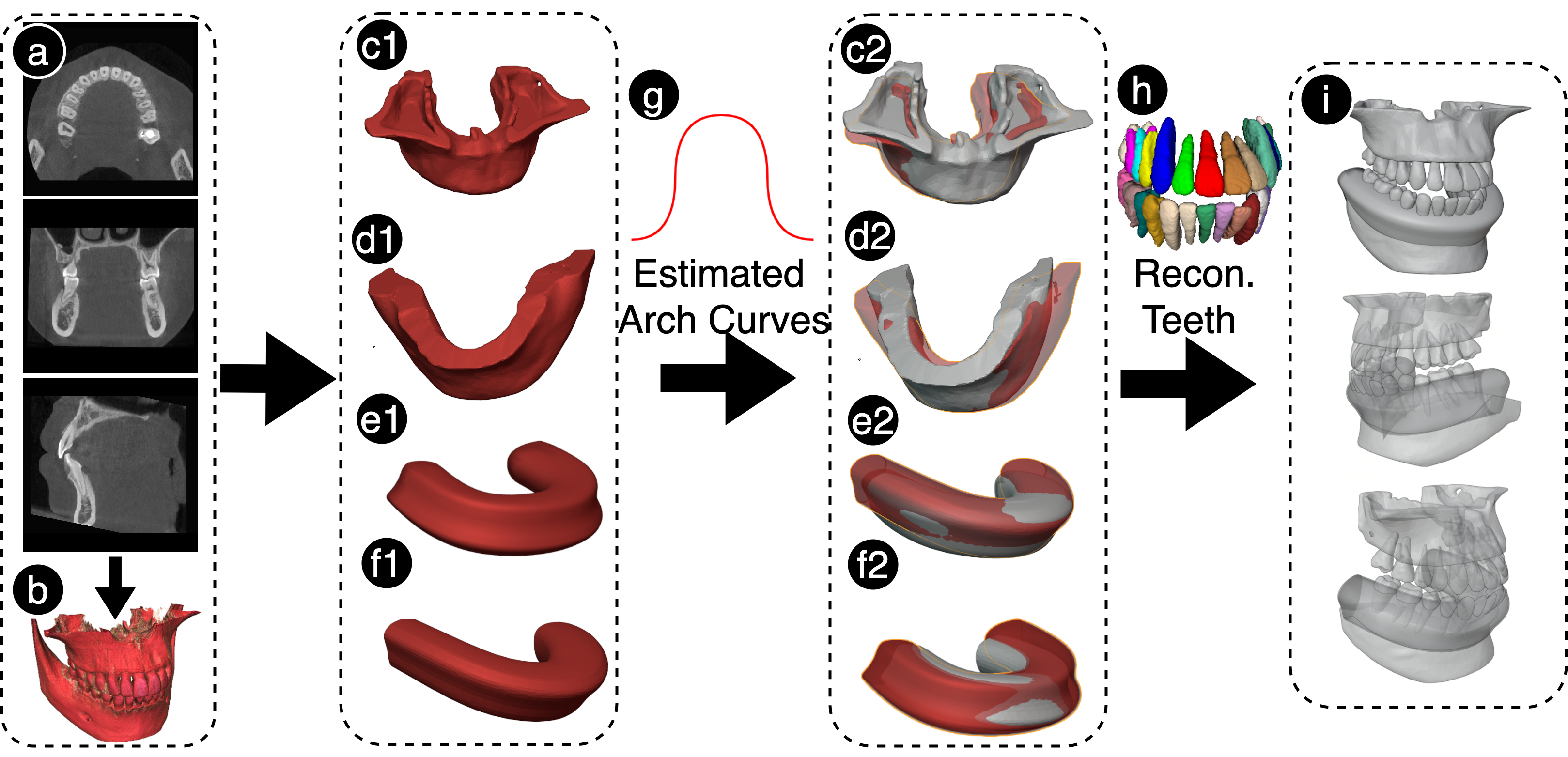}
    \caption{}
    \label{fig_prebuilt}
  \end{subfigure}
  \vspace{-1em}
  \caption{ (a) Teeth naming, numbering and coloring rule. (b) Gum and jaw bone templates building method, and their application in the deployment stage for a complete oral cavity model. 
}
\vspace{-5mm}  
\end{figure}

\paragraph{\textbf{Model Architecture.}}
Our task of teeth reconstruction has two unique challenges from the existing voxel-based work. 
(\textit{i}) The reconstruction contains multiple objects (teeth) rather than a single object as in \cite{tatarchenko2019single,xian2018monocular,choy20163d}. 
(\textit{ii}) The input image of X-ray has a higher resolution than existing work (\textit{e.g.}, 128$\times$128 \cite{choy20163d}), which calls for higher computational and memory efficiency of model.
To tackle both challenges, we decompose the task into two sub-tasks of teeth localization and patch-wise single tooth reconstruction. 
% arch 
Figure \ref{fig_model} shows the overall architecture of our model.
% feature extract 
The input 2D panoramic X-ray $X \in \mathbb{R}^{H \times W}$(Figure \ref{fig_model}(a)), where $H$ and $W$ are image height and width, is fed to a feature extraction subnet (Figure \ref{fig_model}(a)) of a 2D encoder-decoder structure for capturing a deep feature map of the same resolution as the input X-ray. 
% , and is trained to encode contextual information. \xac{what do you mean by contextual} 
% segmentation  
Given the feature map, a segmentation subnet (Figure \ref{fig_model}(b)), which consists of stacked convolutional layers and a $Sigmoid$ activation, maps it into a categorical mask $Y_\text{seg} \in \mathbb{Z}^{H \times W \times K}$, where $K=32$ denotes the maximum number of tooth category. 
Figure \ref{fig_numbering} demonstrates the tooth numbering rule we used by following the World Dental Federation Notion \cite{hickel2010fdi}.  
% bounding box
Moreover, the tooth localization of bounding boxes is further derived from the segmentation map by keeping the largest island per tooth, as shown in Figure \ref{fig_model}(3). 
% single tooth recontruction
According to the tooth localization, a tooth reconstruction subnet (Figure \ref{fig_model}(c)) performs patch-sampling for all teeth from the aforementioned deep feature map (Figure \ref{fig_model}(2)), and back projects them into 3D tooth shapes represented in 3D occupancy map (Figure \ref{fig_model}(4)), using a 2D-encoder-3D-decoder structure similar to \cite{choy20163d}.  
% share a common feautre: two reasons 
Note that the deep feature map (Figure \ref{fig_model}(2)) is shared for both segmentation and tooth reconstruction sub-tasks in order to increase the compactness and generalization of the model. 
% assmebnle flatten 
By assembling the predicted tooth volumes (Figure \ref{fig_model}(4)) according to their estimated localization from X-ray segmentation (Figure \ref{fig_model}(3)), we can achieve a flatten 3D reconstruction of teeth. 
% curve -> whole model 
The flatten reconstruction is then bent to an estimated dental arch curve (Figure \ref{fig_model}(5)) for the final 3D teeth reconstruction as shown in Figure \ref{fig_model}(6)). 
% end to end trianbale, supervised 
The parameters in all the subnets of the model (Figure \ref{fig_model}(a,b,c)) can be optimized in an end-to-end fashion for the optimal performance. 
The training strategy and dataset are described as below. 
% more detail 
More details about the model can be found in Supplementary Material 1.2.
% \ref{supp_model}. 
% \xac{how did you get training data?}

\begin{figure*}
\includegraphics[width=0.85\textwidth]{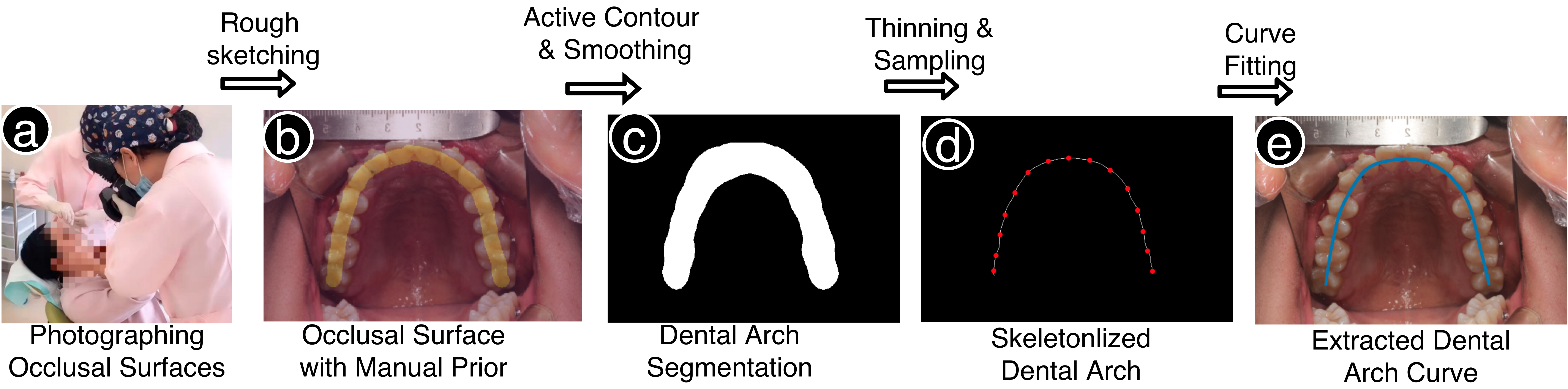}
\caption{Semi-automatic dental arch curve estimation from occlusion photos.} 
\label{fig_extract}
\end{figure*}

\paragraph{\textbf{Unsupervised Dental Arch Curve Estimation.}}
% curve, since it is lost  
The dental arch curve needs to be estimated since such information is lost during the circular rotational screening process of a panoramic X-ray imaging system \cite{haring2000dental}. 
% multiple methods
Multiple methods can be applied for the estimation, \textit{e.g.}, using average shape from general population \cite{muhamad2015curve}, and $\beta-$curve fitting with measured width and depth of oral cavity \cite{braun1998form}. 
% in this work, oral photo 
In this work, we propose a semi-automatic pipeline to accurately extract dental arch curve from occlusal surface photos without supervision, as shown in Figure \ref{fig_extract}. 
% manual labeling, 
First, macro shots of upper and lower occlusal surfaces are taken from a patient (Figure \ref{fig_extract}(a)), and are roughly labeled for teeth regions with simple sketching (Figure \ref{fig_extract}(b). 
% \xac{hard to see the light yellow stroke}).  
% active contour, sampling, 
The dental arch area can then be obtained with the active contour algorithm \cite{williams1992fast} applied on the sketches as priors (Figure \ref{fig_extract}(c)), and further skeletonlized for an initial dental arch curve (Figure \ref{fig_extract}(d)). 
% curve fit 
The final smooth dental arch curve is achieved by fitting a 
cubic curve to the uniformly sampled data points from the initial curve (Figure \ref{fig_extract}(e)). 

\paragraph{\textbf{Complete Oral Cavity Model.}}  
% we build what 
To achieve a complete oral cavity model, \textit{OralViewer} is embedded with a set of pre-built template models for gums and jaw bones. 
% As shown, in the template buildin stage, structure with thresholding methid 
As shown in Figure \ref{fig_prebuilt}, to build the templates, a Cone Beam CT from an adult male was collected \ref{fig_prebuilt}(a), and pre-processed with intensity thresholding \cite{naumovich2015three} for extracting the skull structure \ref{fig_prebuilt}(b).
% extracted, smoothed for XXXXX, 
Then, the upper jaw (Figure \ref{fig_prebuilt}(c1)) and lower jaw (Figure \ref{fig_prebuilt}(d1)) bone models were constructed by removing tooth structures, hole filling, and smoothing,  
% gum is what. 
while the upper gum (Figure \ref{fig_prebuilt}(e1)) and low gum (Figure \ref{fig_prebuilt}(f1)) models were constructed as the smooth volumes embodying the corresponding jaw bones. 
% For each case, we register and deform to apply, as shwon
For the reconstruction for each patient in the deployment stage, the pre-built gum and jaw bone models are first registered and aligned to the estimated dental arch curves (Figure \ref{fig_prebuilt}(g)).
% By asseemnling with XX, we achieve 
Then the deformed gum and jaw bones model (Figure \ref{fig_prebuilt}(c2,d2,e2,f2)) are assembled with the 3D reconstructed teeth for the complete oral cavity model as the example shown in Figure \ref{fig_prebuilt}(i).   
% The whole model sturtcture awareness, for manualy by doctore.  
We expect the averaging models of gum and jaw bone from CT scans of multiple individuals can further improve the reconstruction quality, while the current templates have been shown valid for the surgical demonstration purpose according to dentists as detailed in the Expert Study section. 

% \begin{figure*}
% \includegraphics[width=\textwidth]{samples/result_assemble.pdf}
% \caption{Overall architecture.} 
% \label{fig_extract}
% \end{figure*}

\paragraph{\textbf{Training Dataset.}} 
% we utlized paramcerzied depp cnn for teeth strcutre. 
\textit{OralViewer} utilizes a deep ConvNet for estimating patient-specific 3D teeth structure from a 2D panoramic X-ray.
% intritiveluy, however, issue. 
Intuitively, the model can be trained with the supervision of patients' paired data of 3D teeth structures obtained from the teeth labeling of CT and panoramic X-ray. 
% paired ct for alignment 
However, tooth structures from X-ray and CT are misaligned due to different postures during screening, \textit{e.g.} head directions and occlusion condition.
% , and no existing 2D/3D registration method has been developed to handle the issue \cite{markelj2012review}. 
% to train, we build in0-home dataset. 
As such, we propose to collect CT scans, and synthesize their corresponding panoramic X-rays as the model input.
The synthesis is valid since the CT scans contain  full 3D information of oral cavity, while panoramic radiographs are the 2D projections of them. 
% synthesised -> shown in graph + augmentation 
Several previous work has demonstrated promising results for high quality X-ray synthesis from CT, and we employed the method from Yun \textit{et al.} \cite{luo2016automatic} in this work. 
% in total 
In total, our in-house dataset contains 23 pairs of 3D CT scans and synthesized panoramic X-rays.
% , each with a pixel spacing from 0.250mm to 0.434mm.
Moreover, another 39 real panoramic X-ray scans have also been included for augmenting the training process of the segmentation subnet of the model. 
All CBCT scans and panoramic X-rays have been first labeled with pixel-wise tooth masks by 3 research team members, and then reviewed by 2 board-certificated dentists.
We randomly split the paired data into 21 scans for training, and 2 scans as the validation set for determining the early stopping of the training process. 

\paragraph{\textbf{Training Strategy.}} 
% two stages 
We employed a two-stage training paradigm.
% a + b for location.
In the first stage, we train the deep feature extraction subnet and segmentation subnet (Figure \ref{fig_model}(a,b)) for tooth localization, which servers as the starting point for the training of tooth reconstruction subnet since it requires tooth patch sampling ability.  
% loss 
For the segmentation loss $L_\text{seg}$, we define it to be the average of dice loss across all tooth categories. 
% by considering that a pixel on X-rays can be of multiple categories with teeth overlaps. 
% a + b + c for performance 
In the second stage, we train the whole model including tooth reconstruction subnet (Figure \ref{fig_model}(c)) for the optimal performance of both tooth localization and reconstruction, with the loss defined as the sum of the aforementioned segmentation loss $L_\text{seg}$ and a reconstruction loss $L_\text{recon}$. 
The $L_\text{recon}$ was set as the 3D dice loss between the ground-truth volume and the predicted volume.  
% We employed intensive image augmentation to increase the model generalization, which includes image shifting, rotation, scaling, bias correction \cite{tustison2010n4itk} and adding Gaussian noise. 
% Finally, we implemented our model in Pytorch, and trained the model for the experiments on three NVidia Titan Xp GPUs. 
More details about model training can be found in Supplementary Material 1.3. 
% \ref{supp_training}. 
% % 5. 1 image + 20 patches on GPU with 3 GPU 
% For each GPU, we set the batch size of panoramic radiograph as 1, and the batch size of tooth patches as 10.
% % 6. augmentation: shift/noise/etc for reduce overfitting. 
% Besides, standard augmentations are employed for images, including random shifting, scaling, rotating and adding Gaussian noise. 
% 7. implementation. 

\subsection{Demonstrating Surgery with Virtual Dental Instruments} 
% tools descptions (R3)
\textit{OralViewer} provides a web-based 3D demonstration tool for dentists to demonstrate surgery steps on a patient's oral cavity model with virtual dental instruments. 
% implementation for surgies
The dental instruments allow dentists to express what and where an action is applied to the oral cavity, and demonstrate the effect on the model in real-time (\textbf{R3}).
Moreover, dentists can use simple sliders to customize the
animation effect of the instruments to better suit their particular needs and preferences. 
% according to dentists steps: : XXX
By discussing with the dentists, the current tool's implementation consists of six common dental instruments
% for enabling the demonstration of crown lengthening and apicoectomy\xac{the first time i read it, i missed that these are the only two surgeries we cover; need to describe what they do, justify why choosing these two and only these two.}
: (\textit{1}) surgical scalpel, (\textit{2}) fissure bur, (\textit{3}) handpiece, (\textit{4}) syringe, (\textit{5}) curette, 
% (\textit{6}) dental membrane, 
and (\textit{6}) artifacts. 
% \xac{do these cover most dental operations? show a figure of the instruments?}
% apply to others 
% Moreover, we expect the demonstration tool can be easily extended for other surgeries by implementing more virtual dental instruments following our method. 
% \xac{do we cover methods to create such virtual tools?}
% % we first, 
In this section, we first show the overall workflow of using the demonstration tool.  
% we show method for each tool 
We then describe the technical details of each virtual dental instrument, followed by the tool implementation. 
% % details about surgeries can be refered to 

\paragraph{\textbf{Overall Workflow.}}
\begin{figure*}
\includegraphics[width=\textwidth]{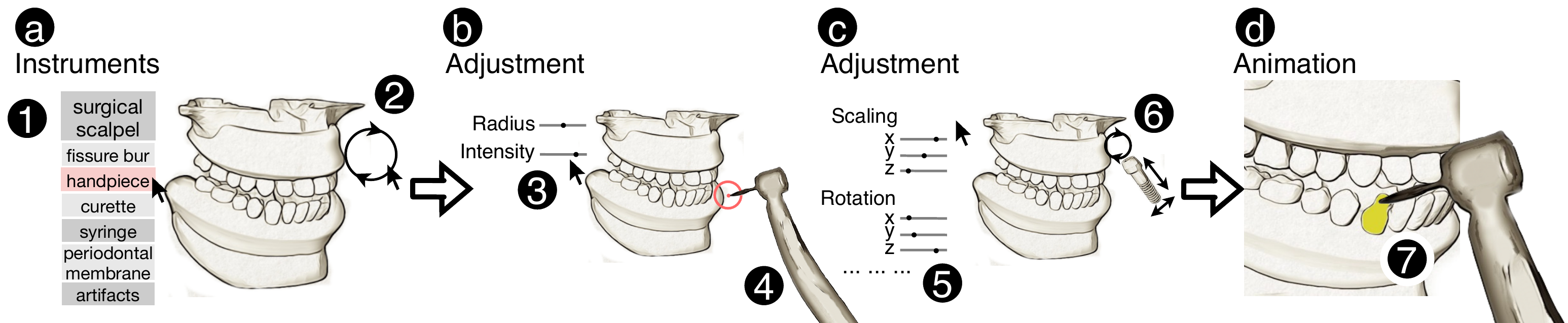}
\caption{Overall workflow for demonstrating dental surgeries.} 
\label{fig_tool}
\end{figure*}

% Figure \ref{fig_tool} shows the overall workflow using the demonstration tool. 
% \xac{which tool? there are 7?}. 
% loading 
As shown in Figure \ref{fig_tool}(a), a dentist start with importing a reconstructed 3D oral cavity model generated from the aforementioned pipeline (Figure \ref{fig_tool}(1)), which can be viewed freely with rotation and scaling.
% tools 
To apply a virtual dental instrument, the dentist selects the instrument from a list (Figure \ref{fig_tool}(2)). 
% artifacts 
% adjusts
Upon the selection, the corresponding instrument model (Figure \ref{fig_tool}(4,6)) is visualized, and can be controlled by using the mouse to move and operate on the oral cavity model. 
For instruments \textit{1-5},  
% \xac{what does this mean? maybe define it at the beginning of this subsec} 
\eg scalpel as shown in Figure \ref{fig_tool}(b), their animation effect on dental structures can be customized by changing a set of parameters (Figure \ref{fig_tool}(3)); 
while for dental artifacts, \eg implant as shown in Figure \ref{fig_tool}(c), their shapes and directions can also be adjusted to tailor for the patient's condition (Figure \ref{fig_tool}(5)).
% apply 
The selected instrument can be directly applied to a dental structure for demonstrating effects with clicking, pressing, and dragging (Figure \ref{fig_tool}(d)).
The oral cavity model can also be freely rotated by press-and-drag and scaled with wheel to adjust an optimal view for the demonstration and manipulation.
The effect can be dynamically reflected on the structure in real-time, with the operated structure highlighted for visualization (Figure \ref{fig_tool}(7)).
% repeative to finish 
A typical dental surgery consists of several sequential steps using multiple dental instruments, which can be demonstrated on a 3D oral cavity model by operating with each dental instrument following the aforementioned steps of instrument selection (Figure \ref{fig_tool}(a)), adjusting (Figure \ref{fig_tool}(b,d)), and animating (Figure \ref{fig_tool}(d)). 
% \xac{suggest replacing all 'animate' words with 'demonstrate'?}
% \xac{instead of saying what the tool can generally do, maybe describe a specific procedure/operation as an example using fig 5?}

% what 
% how use in surgery 
% how use 
% how use 
% hwo use 
% #1 knife (cut and back)
\paragraph{\textbf{\#1 Surgical Scalpel \& \#2 Fissure Bur.}}
% \begin{figure*}
% \includegraphics[width=\textwidth]{samples/tool_knife.png}
% \caption{Overall architecture.
% \xac{need to zoom into the operated region}
% } 
% \label{fig_tool_knife}
% \end{figure*}
\begin{figure*}
\includegraphics[width=\textwidth]{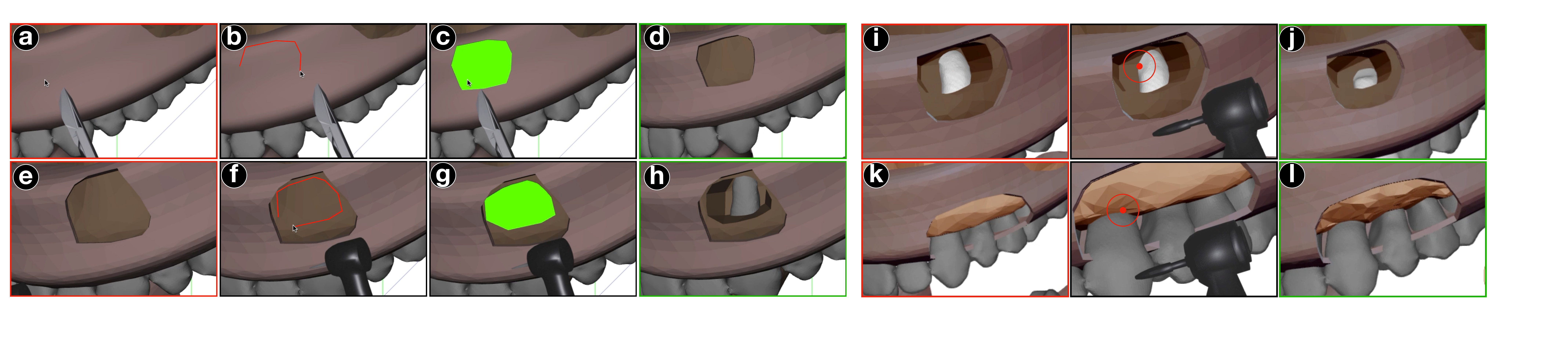}
\caption{Example use cases of virtual surgical scalpel (top left), fissure bur (bottom left), handpiece for tooth tip resection (top right) and jaw bone lowering (bottom right).} 
\label{fig_tool_knife}
\end{figure*}
Both tools are used to incise dental structures: surgical scalpel can be applied to gum tissue for the creation of periodontal flap; while fissure bur can create holes on jaw bones for the exposure of the root tip. 
To use the tools, a dentist describes the desired incision location and its size by creating a region from a sequence of mouse clicks, where the created boundary is visualized with red lines (Figure \ref{fig_tool_knife}(b,f)). 
Upon a closed-loop boundary is formed, the incised region is highlighted in green (Figure \ref{fig_tool_knife}(c,g)), and the corresponding part of the dental structure is then removed upon the dentist's confirmation. 
The design allows dentists to perform any type of incision according to a patient's condition, \textit{e.g.,} semilunar and triangular types of periodontal flap. 
Figure \ref{fig_tool_knife}(a$\rightarrow$d) shows a periodontal flap example using surgical scalpel, and Figure \ref{fig_tool_knife}(e$\rightarrow$h) shows the creation of jaw bone opening based on a flapped gum. 

\paragraph{\textbf{\#3 Handpiece.}}
% \begin{figure*}
% \includegraphics[width=0.6\textwidth]{samples/new_tool_grind.pdf} 
% \caption{Example of use cases of handpiece for tooth tip resection (top row) and jaw bone lowering (bottom row).}
% \xac{what's the difference between images 2-4 in row 1? }
% \xac{for these figures, maybe highlight 1) the action of the virtual tool (\eg movement direction) and 2) the difference of the tooth before and after the operation.}
% } 
% \label{fig_tool_larger_drill}
% \end{figure*}
Handpiece is widely used for shaping bone structures, \textit{e.g.,} tooth and jaw bones. 
A dentist can move the virtual handpiece as mouse cursor to any desired location of grinding.
The grinding effect takes place once with a mouse click, or continuously by pressing and dragging the mouse.
The size and intensity of the grinding effect can be customized using sliders. 
For example, Figure \ref{fig_tool_knife}(i$\rightarrow$j) shows the resection of an exposed root tip in an apicoectomy; 
while Figure \ref{fig_tool_knife}(k$\rightarrow$l) demonstrates it is applied to reduce the upper jaw bone height in a crown lengthening surgery. 
% For example, Figure \ref{fig_tool_larger_drill}(a-b) shows the resection of an exposed root tip in an apicoectomy; 
% Figure \ref{fig_tool_larger_drill}(c-d) shows a tooth is prepared by shaping the room for an artifact crown to sit in, which can happen in a crown lengthening surgery teeth with broken crown but complete root; 
% moreover, Figure \ref{fig_tool_larger_drill}(e-f) demonstrates that handpiece is used to reduce the upper jaw bone height in a crown lengthening surgery, where the tooth cavity below the jaw bone architecture can be revealed. 

% # 5 syringe 
\paragraph{\textbf{\#4 Syringe \& \#5 Curette.}}
\begin{figure*}
\includegraphics[width=0.85\textwidth]{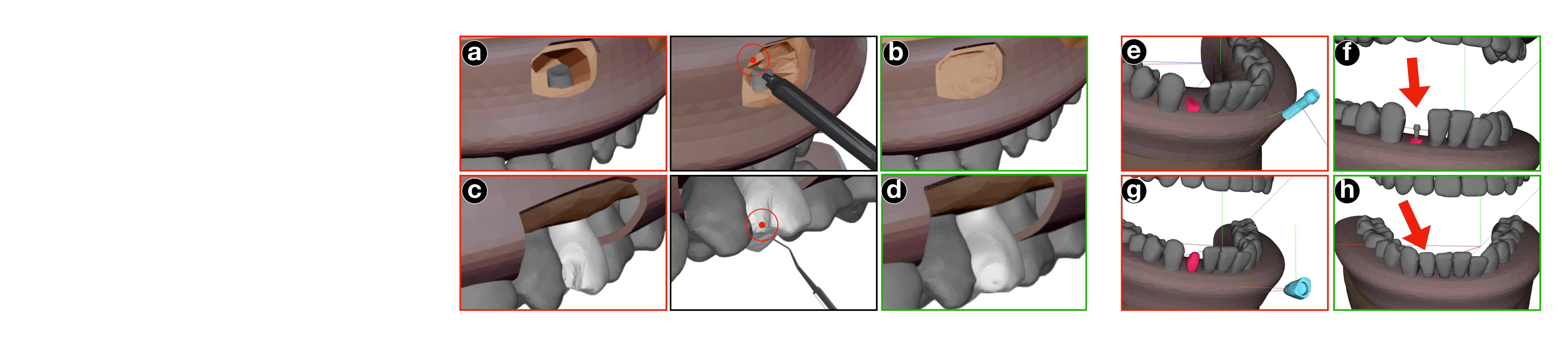}
\caption{Example use cases of syringe for injecting grafting materials (top left), curette for restoring fractured crown (bottom left), applying an artifact implant (top right), and cementing an artifact crown (bottom right).  
% \xac{this fig is clearer than the others in terms of showing the sequence of action and what are the corresponding changes on the teeth}
} 
\label{fig_tool_syringe}
\end{figure*}
Both instruments can be used to fill materials into holes of bone structures, \eg teeth and jaw bones. 
Similarly, a dentist can move either instrument as a cursor to the desired location on the tooth or jaw bone structures and click or press-and-drag for the filling effect. 
The size and intensity of the filling effect can be customized using sliders. 
For example, Figure \ref{fig_tool_syringe}(a$\rightarrow$b) shows the syringe is used to inject bone grafting materials to fill the jaw bone opening, and Figure \ref{fig_tool_syringe}(c$\rightarrow$d) demonstrates a curette is applied on a fractured crown for restoration. 

\paragraph{\textbf{\#6 Artifacts.}}
% \begin{figure*}
% \includegraphics[width=\textwidth]{samples/tool_artifact.jpg}
% \caption{Overall architecture.
% \xac{too small}} 
% \label{fig_tool_artifact}
% \end{figure*}
Artifact crown and implant are included in the demonstration tools. 
To apply them, a dentist starts with importing a pre-defined artifact model, followed by specifying the dental structure of the artifact to be applied on. 
The artifact model and the dental structure are visualized in blue and red (Figure \ref{fig_tool_syringe}(e,g)) 
% \xac{figure?} 
for assisting the operations and demonstration. 
The 3D location of the artifact can be adapted with dragging in 2D viewing planes 
% \xac{how can you drag in 3d?}
, while both the orientation and size
% \xac{changed it to this. plz check.} 
of the artifact can be modified to match the patient's condition using sliders from \textit{X}, \textit{Y}, and \textit{Z} axis. 
Once the artifact has been customized with confirmation from dentists, the artifact model and the operated dental structure are merged as one object.
Figure \ref{fig_tool_syringe} (e$\rightarrow$f, g$\rightarrow$h) shows the cementation of an implant on a resected root and a dental crown on a prepared tooth, respectively, as indicated with red arrows.
% \xac{highlight the crown in h}

% implementation 
\paragraph{\textbf{Implementation.}}
The demonstration tool of \textit{OralViewer} was implemented using OpenGL, JavaScript and three.js. 
The tool can run readily inside a modern browser. 
The effect of gum/jaw bone incision with surgical scalpel/fissure bur and artifact implanting was implemented with the Constructive Solid Geometry (CSG) operations\footnote{\url{https://github.com/oathihs/ThreeCSG}} between: (\textit{i}) generated 3D convex geometry from the input trajectories or pre-built artifact models, and (\textit{ii}) corresponding dental structures. 
The effect of shaping/filling with handpiece/syringe/curette was implemented with mesh sculpting operations\footnote{\url{https://stephaneginier.com/sculptgl/}} including flattening, filling, and scraping. 

% \xac{if possible, move the figure closer to where it is described to make it easier to read.}

\section{Evaluation}
We conducted \one a technical evaluation of the reconstruction pipeline for generating a 3D model of a patient's oral cavity, \two a patient study with 12 participants, and \three an expert study with three board-certified dentists. 

\subsection{Technical Evaluation} 
\paragraph{\textbf{Dataset.}} 
% To assess the accuracy of the reconstruction model, w
We built an in-house testing dataset collected from 10 patients from a local orthodontics hospital.
Each patient was screened for both panoramic X-ray and Cone Beam CT. 
Moreover, since our reconstruction pipeline estimates patient-specific dental arch curves, two photos of occlusion macro photos (for upper and low jaws respectively) of each patient were captured by dentists in the clinic.
To quantitatively evaluate the reconstruction accuracy, patients' tooth structures were manually labeled from CT scans
% The labeling was done 
by two research team members and reviewed by one dentist.
% In total, the dataset consists of 10 paired data of X-ray, CT, 2D tooth mask, 3D tooth structures, and two occlusion photos for upper and low jaw. 

\paragraph{\textbf{Results \& Analysis.}} 

% \begin{figure*}
% \includegraphics[width=0.7\textwidth]{samples/results_acc.pdf}
% \caption{Overall architecture.} 
% \label{fig_results_acc}
% \end{figure*}

\begin{figure}[t]
  \centering
  \begin{subfigure}{.7\textwidth}
    \includegraphics[width=1\linewidth]{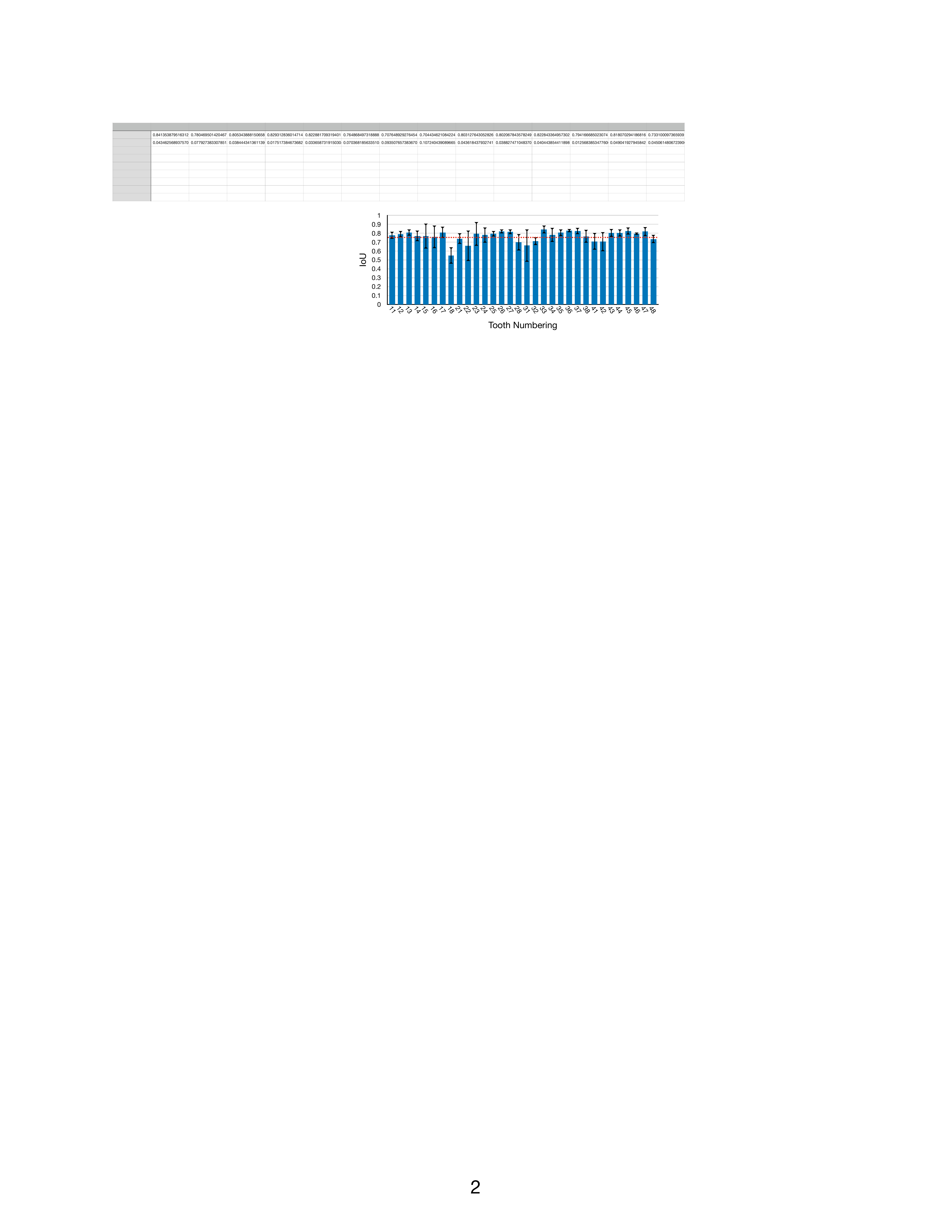}
    \caption{
    }\label{fig_results_acc}
  \end{subfigure}%
  \hfill
  \begin{subfigure}{.25\textwidth}
    \includegraphics[width=1\linewidth]{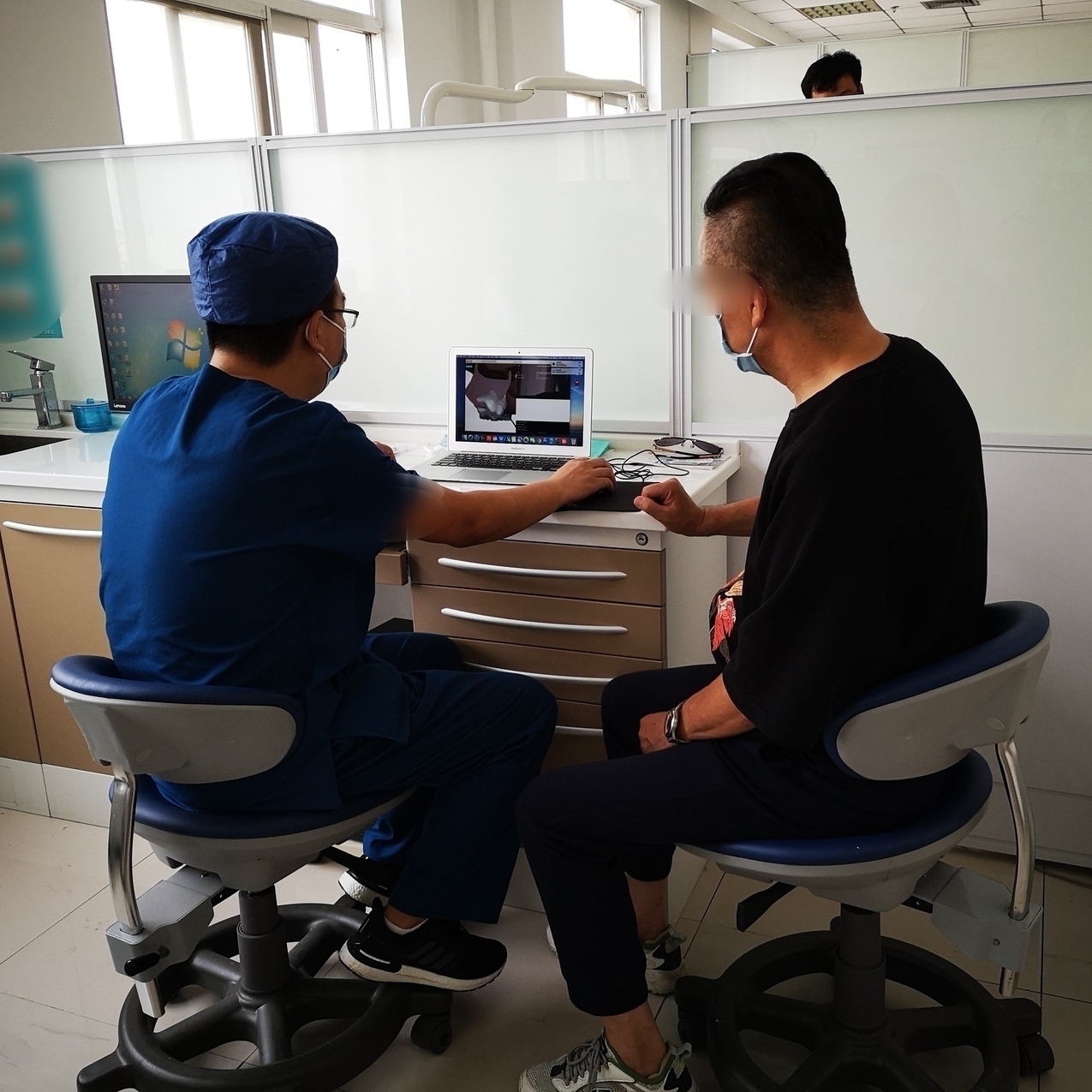}
    \caption{}
    \label{fig_study_example}
  \end{subfigure}
  \caption{ (a): 3D teeth reconstruction accuracy. (b) A participant was demonstrated a dental surgery using \textit{OralViewer} by a dentist.
}
\end{figure}

We applied our reconstruction pipeline with the panoramic X-ray and occlusion photos as input to generate the complete 3D oral cavity model.
We first evaluate the teeth reconstruction accuracy for reflecting patient-specific condition by quantitatively comparing the generated model with labeled 3D structure from CT.
We used the Intersection over Union (IoU) between the predicted volume $P$ and the ground-truth volume $G$ of each tooth as the performance metric, where $IoU = |P \cap G| / (|P \cup G|)$, and $|.|$ denotes the cardinality of a voxel set.
To overcome the misalignment of teeth between X-ray and CT caused by patients' gestures, the upper and lower teeth labeling from CT (3D) was rigidly aligned (using ANTs \cite{avants2009advanced}) to the corresponding part of the reconstructed 3D teeth (3D) from the model before the IoU calculation. 
% \xac{how to align the two models in 3d space in terms of position and orientation?}
Figure \ref{fig_results_acc} reveals the IoUs for all the 32 categories of tooth, where our model achieves a mean IoU of 0.771, with patient-wise $std=0.031$ and tooth category-wise $std=0.062$. 
% \xac{how about participant-wise average?}
The result indicates the effectiveness of our method for reconstructing 3D teeth model from 2D x-ray image. 
Yet, we also find that the method has a consistent lower accuracy for wisdom teeth (numbering 18, 28, 38, and 48) than the others, which can be caused by: (\textit{i}) the shape of wisdom teeth is more variant across different persons, and (\textit{ii}) relatively fewer samples are available for ConvNet training (16 out of 23 cases in the training dataset containing one or more wisdom teeth). 
% More results on the evaluation of the segmentation sub-task can be found in Supplementary Material \ref{supp_evaluation}. 

\begin{figure*}
\includegraphics[width=\textwidth]{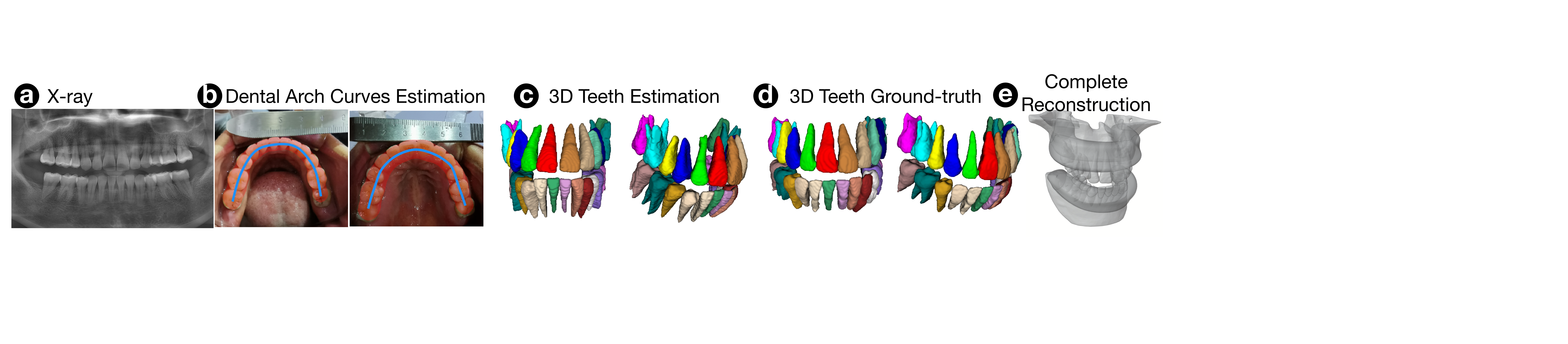}
\caption{Reconstruction example from testing dataset.} 
\label{fig_demo_case}
\end{figure*}

We then qualitatively evaluate the 3D reconstruction of the complete oral cavity model. 
Figure \ref{fig_demo_case} visualizes an example case from the testing dataset: 
% Figure \ref{fig_demo_case}(a) shows 
the input panoramic X-ray (a), 
% Figure \ref{fig_demo_case}(b) show 
the semi-automatic dental arch extraction results for both upper and lower jaws (b),
% Figure \ref{fig_demo_case}(c) visualizes 
the reconstructed 3D teeth model from the 2D X-ray and the estimated dental arch curves (c), 
% while Figure \ref{fig_demo_case}(d) shows 
the ground-truth teeth structures extracted from the CT image (d), and 
% Figure \ref{fig_demo_case}(e) demonstrates 
the complete oral cavity model with the pre-defined jaw bone and gum models registered and assembled with the reconstructed teeth (e). 
Note that the shown case has 24 teeth, rather than 28 teeth as a normal adult. This is possibly because 4 teeth have been extracted during a previous orthodontic operation. 
The reconstructed results clearly reflect this individual's condition, and show the effectiveness of our reconstruction pipeline for generating 3D oral cavity models. 
% \xac{marked as unread}

\subsection{Patient Study}
\textit{OralViewer} enables a dentist to help laymen patients understand procedures of dental surgeries by 3D demonstration.
To validate the feasibility of our approach, we investigate the research question concerning the effect of \textit{OralViewer}:

\begin{itemize}
    \item \insec{RQ1: can \textit{OralViewer} improve patients' understanding of a surgery}? 
\end{itemize}
% \xac{these three sub-questions sound like the same thing to me}

\subsubsection{Participants} We recruited 12 patients from the clinic of an orthodontics hospital  (4 females and 8 males, aged between 21 to 64 years).
Each participant was demonstrated with one surgery of crown lengthening or apicoectomy. 
None of the participants had received the dental surgery before. 
% \xac{do we know if they had such surgery before?}
% Although \textit{OralViewer} is intended for users actually undergoing a surgery, 
Note that due to limited patient resources, we were only able to recruit participants who came in for regular dental check-up but did not actually undergo such dental surgeries. 
% In the future we plan to test \textit{OralViewer} on patients who need to receive such surgeries and further collect their post-op feedback.
% This participant pool selection was mainly due to the difficulties in recruiting with the populations within a limited time. 
%below may place somewhere else 
The detailed demographic information of our participants can be found in Table 1 from Supplementary Material 1.4.
% \ref{supp_users}. 

\subsubsection{Procedure} The patient study consists of the following
key activities: 

{\it Tutoring dentists how to use the system.}
To clinically validate \textit{OralViewer}, we collaborated with three board-certified dentists (E1: male, 25 years of practice; E2: male, 17 years of practice, E3: female, 11 years of practice) for carrying out the surgery demonstrations. 
We first introduced \textit{OralViewer}, let each dentist follow a step-by-step onboarding tutorial, and answered their questions about utilizing the program. 
Then dentists were free to continue trying out \textit{OralViewer}'s virtual dental instruments until they felt they were able to use the system independently.

{\it In-clinic study.}  
In order to compare \textit{OralViewer} with current patient education method, we randomly split the participants into an experiment group of 7, which was demonstrated \textit{OralViewer}, and a control group of 5, which was demonstrated an X-ray and verbal descriptions as per the dentists' regular practice. 
% based on the dentist's current practice. \xac{what exactly did they use in control group?}
% using any method according to a dentist's developed habit. 
Participants were randomly assigned one of the two dental surgeries, \eg apicoectomy or crown lengthening, to receive a demonstration: 4 participants in the experiment group and 3 in the control group were demonstrated apicoectomy; while 3 in the experiment group and 2 in the control group were demonstrated for crown lengthening. 
Each study happened in one of the three dentists' clinics and \textit{OralViewer} was accessed as a web app using dentists' own computers (Figure \ref{fig_study_example})
% \xac{refer to figure}
. 
Details on the surgery and dentist assignment for each participant can be found in Table 1 from Supplementary Material 1.4.
% \ref{supp_users}. 

{\it Exit interview.} 
After the explanation, we interviewed each participant to verbally describe the surgery procedures by focusing on: (\textit{i}) what steps are included, and (\textit{ii}) how (by using what instrument) a procedure is applied. 
Their answers were recorded and later compiled for evaluating their understanding of surgeries. 

\subsubsection{Analysis \& Results}
\begin{figure*}
\includegraphics[width=0.8\textwidth]{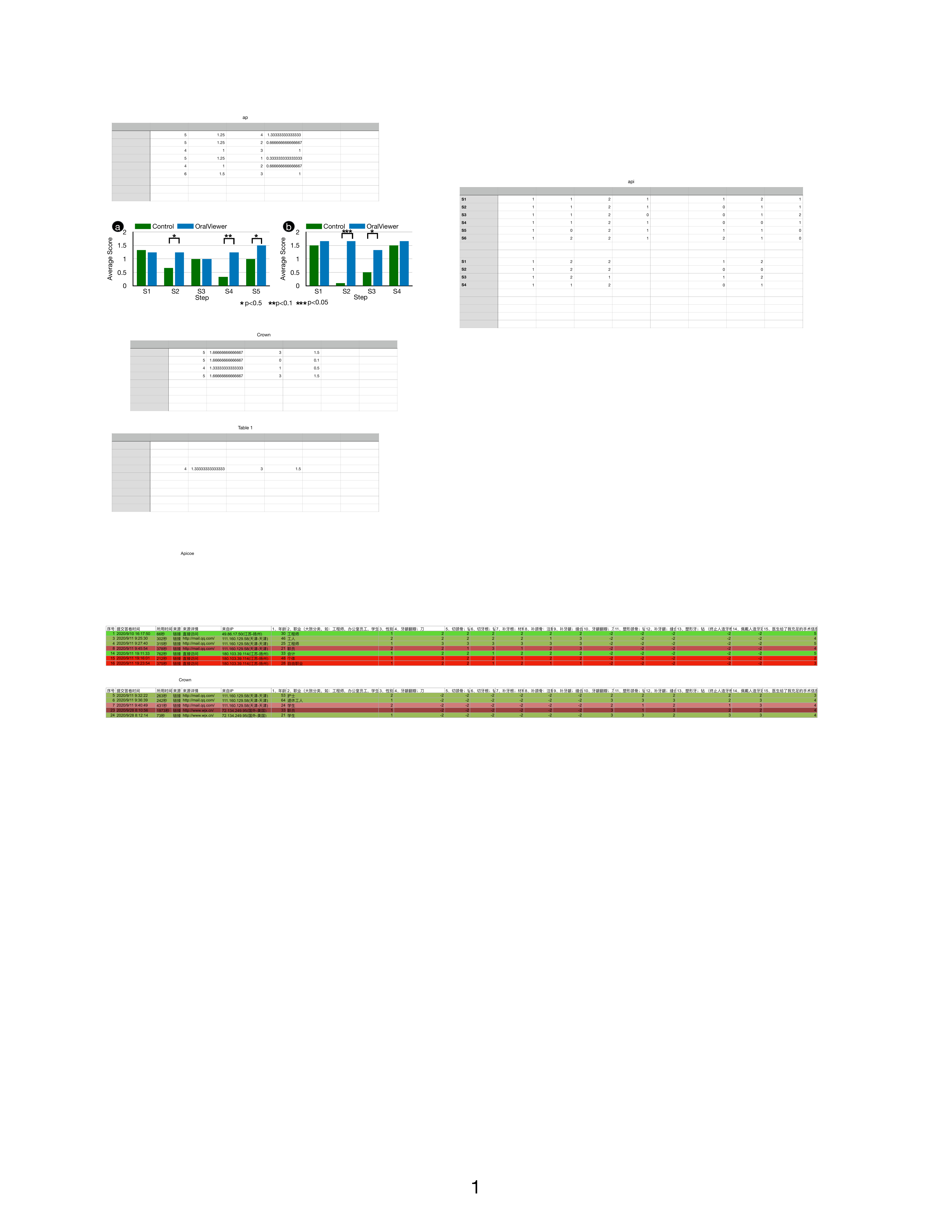}
\caption{Participants' understanding score distribution for (a) apicoectomy and (b) crown lengthening. S\textit{x} represents key steps within the two surgeries. 
% \xac{use 'OralViewer' instead of 'Experiment'; explain what 'Sx' means as it might be confused with subjects}
% \xac{should be *** p<0.05}
} 
\label{fig_results_user}
\end{figure*}

We scored the participants' understanding on a surgery based on whether the key steps of the surgery were described in their answers. 
% \xac{i replaced 'procedure' with 'step' because procedure is similar to surgery.}
% \xac{why is this the right way to evaluate their understanding of the procedure? can we expect a patient to learn these specialized words so quickly?}
Specifically, there are five key steps for apicoectomy: (S1) periodontal flap, (S2) jaw bone opening, (S3) root tip removal, (S4) root tip sealing, and (S5) grafting material injection; 
% \xac{where is S5?} 
while five key procedures were considered for crown lengthening: (S1) periodontal flap, (S2) jaw bone shaping, (S3) tooth preparing, and (S4) artifact implanting. 
An answer regarding a step is scored as: \textit{0} if the step was not described, \textit{1} if the step was described but the applied dental instrument was not, and \textit{2} if both the step and its dental instrument were mentioned or described.
Note that the exact names of step/instrument were not required to be mentioned in an answer for a score --- a step/instrument was counted if it was described or indicated from a patient's answer.   
% \xac{i used 'mentioned' for naming the step/instrument exactly and 'described' to mean referring to them but maybe not in the exact tecnical words.} 
Figure \ref{fig_results_user} shows the average score for each step within the experiment and control group for the apicoectomy (Figure \ref{fig_results_user}(a)) and crown lengthening (Figure \ref{fig_results_user}(c)). 
% \xac{perform a t-test to meausre significance. still, reviewers might ask.
% you should indicate significance in fig 10 like this example \url{https://live.staticflickr.com/65535/50313900571_c951d42f0e_h.jpg}
% }
We can see that \textit{OralViewer} significantly improves patients' understanding in three out of the five steps for apicoectomy and two out of the four steps in crown lengthening, while the improvement in the other steps is not statistically significant. 
% for 8 out of 10 steps by having a gaining participants higher scores than the control group. 
Moreover, the overall average score among all steps between the experiment and control group was 1.36 \textit{vs.} 0.85, which also indicates \textit{OralViewer} can significantly improve the patients' understanding with $p<0.05$. 
% \xac{is the difference significant?}

\subsection{Expert Study}
To clinically validate \textit{OralViewer}, we interviewed 3 dentists from the patient study after they finished all the assigned patient demonstrations.
Each dentist had done at least one demonstration using \textit{OralViewer} for both types surgeries (E1: 3 times; E2: 2 times; E3 2 times. Details see Supplementary Material 1.4). 
% \ref{supp_users}).
% \xac{be specific, who did how many on which surgery?}
We investigate the following questions: 
\begin{itemize}
    \item \insec{RQ2: Usability ---  do dentists have difficulty using virtual dental instruments in \textit{OralViewer}?}  
    \item \insec{RQ3: Validity --- is \textit{OralViewer}'s demonstration effect clinically valid for patient education?} 
    \item \insec{RQ4: Preference --- do dentists prefer a system like \textit{OralViewer} as a tool for performing patient education for surgeries?} 
\end{itemize}

% To answer RQ2 and RQ3, 
We asked each dentist to rate their agreement (from 1-strongly disagree to 7-strongly agree) with statements about the usability and demonstration effect of the instruments and the resultant 
% Moreover, such agreement rating was also made for the 
demonstration effect of 3D oral cavity models (RQ2 \& 3), as well as their preference (RQ4). 
% Following each rating, we asked experts to comment on what factors led to their rating.
% To answer RQ4, we asked each expert to rate agreement (from 1-strongly disagree to 7-strongly agree) with two preference statements, and inquired for the reasons of the scores. 

% \subsubsection{Analysis \& Results}

\begin{figure*}
\includegraphics[width=0.75\textwidth]{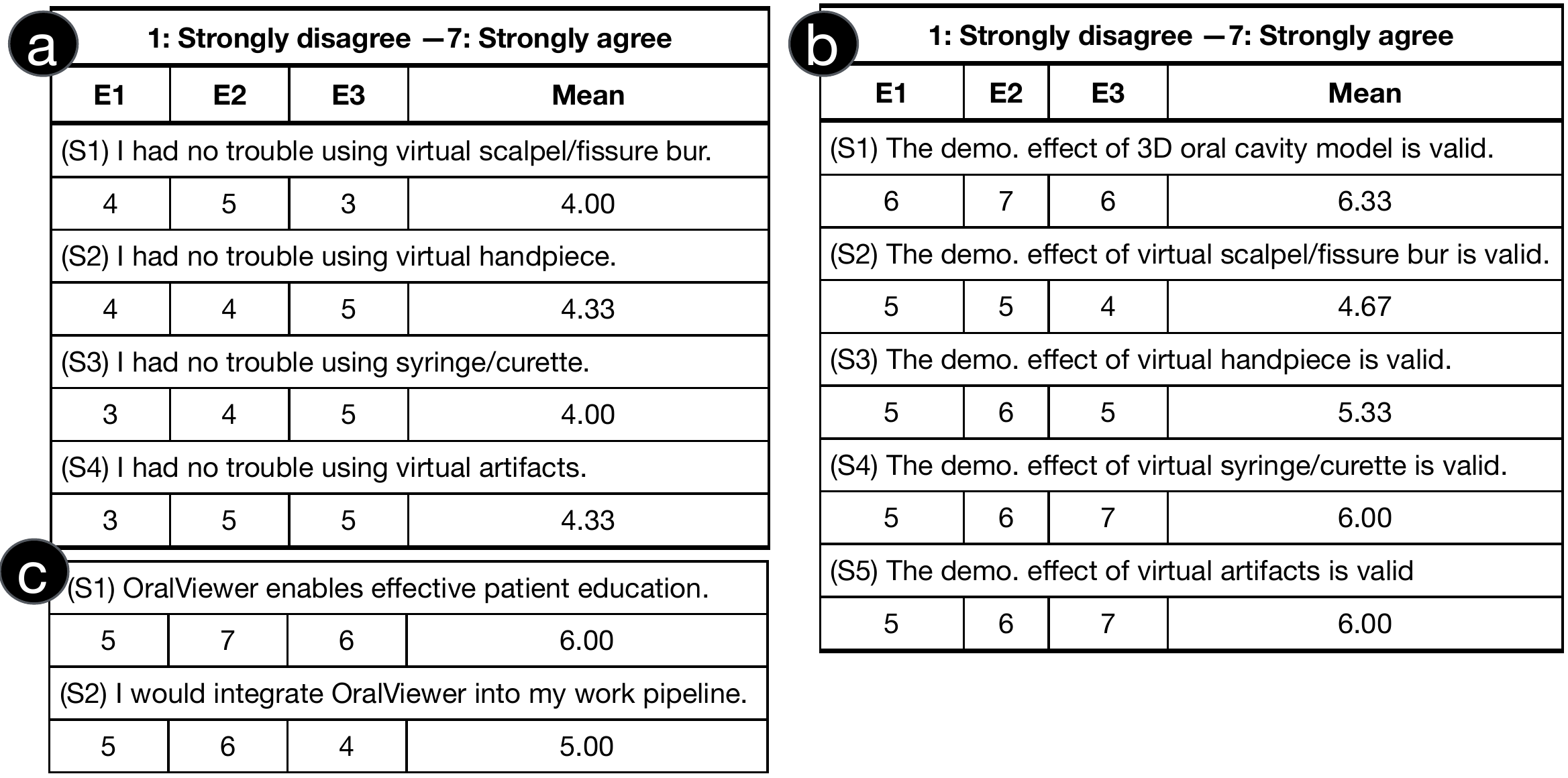}
\caption{Experts' scores for \textit{OralViewer} about (a) usability, (b) demonstration effect validity, and (c) preference. 
% \xac{stack c under a so the entire figure is more compact; ``i like ...'' seems not very professional and not specific to patient education.
% for the first question in c, remove 'more': try not to ask about subjective comparison because the participant is often biased towards new technique you introduce to them.
% }
} 
\label{fig_results_expert}
\end{figure*} 

\subsubsection{Usability}   
We measured the usability for the four types of virtual instruments involved in the demonstration of apicoectomy and crown lengthening: (S1) surgical scalpel and fissure bur for gum/jaw incision, (S2) handpiece for jaw/tooth shaping, (S3) syringe and curette for tooth filling/material injection, and (S4) dental artifacts for implanting. 
% \xac{the 'S' here is confusing with the 'S' used to describe different steps}
Figure \ref{fig_results_expert}(a) shows the questions and experts’ scores.
While all the experts successfully carried out all the demonstrations with patients using \textit{OralViewer}, a major issue raised by experts was that the virtual instrument control with mouse was unfamiliar to dentists: 
% First, \textit{undo} and \textit{redo} functions for each dentist's operating should be added into \textit{OralViewer} (E1,E3).  
% \xac{undo and redo are so basic and trivial in hci papers, people usually don't mention it. that is, everyone knows an interactive system, no matter what it does, needs to have undo/redo. maybe don't mention this part. it will appear a little naive to the reviewers.}
% On one hand, dentists might need \textit{undo} to revoke a misoperation; on the other, \textit{undo} and \textit{redo} can help dentist's demonstrating the change of an operation for procedure explanation. 
% \textit{E.g.}, as mentioned by E3, she would like to use such functions to illustrate the height changes of jaw bones before and after grinding.
it is different from the way dentists using real dental instruments in a surgery, which can lead to a steep learning curve (E2,E3). 
E3 suggested that an implementation of \textit{OralViewer} on touch screen, \eg iPad, with the control of virtual instruments using stylus should be more intuitive to dentists. 

\subsubsection{Validity}
The experts rated the demonstration effect of \textit{OralViewer} for the 3D oral cavity model (S1) and the four types of virtual instruments (S2-S5).
Figure \ref{fig_results_expert}(b) shows the questions and experts' ratings, where \textit{OralViewer} achieves a mean score of 5.67 out of 7.
All experts agreed that our reconstructed oral cavity model and virtual instruments are clinically valid for patient education. 
Regarding the oral cavity model, experts confirmed that it contributes to the surgical demonstration because (\textit{i}) patients are able to see structures that they cannot observe from a mirror, \eg molar teeth, using rotation in 3D (E1, E2, E3); and (\textit{ii}) the patient-specific teeth 
% \xac{did we use teeth models specific to the 12 participants? if so, mention it.} 
can not only let patients understand their conditions better (E2, E3) but also raise their interests in learning more about operating on such conditions (E3).
Besides, E3 suggested that oral cavity model can be improved by modeling root canals within tooth (detail in the Discussion section).
Regarding virtual instruments, experts agreed that they are valid for the patient education purpose (E1, E2, E3), and preferred the visualization of the instruments, which can help patients comprehensively understand what to expect during a surgery (E3). 
Moreover, experts also suggested that the appearance of virtual instruments can be dynamic altered upon users customizing their effect to further improve the visualization (detailed in the Discussion section). 

% Moreover, although the virtual instruments did not precisely reflect the effects as in real surgeries, \eg 
% \xac{what is the imprecise part? example?}
% , they are valid for the patient education purpose (E1,E2,E3). 

\subsubsection{Preference} 
We asked experts about their preference of \textit{OralViewer} from two perspectives: (S1) \textit{OralViewer} enables effective patient education and (S2) I would integrate \textit{OralViewer} into my existing practice. As shown in Figure \ref{fig_results_expert}(c), the experts agreed that \textit{OralViewer} enables effective patient education with a mean score of 6.00 out of 7. 
% Moreover, E2 noted that \textit{OralViewer}'s visualization   that the visualization of \textit{OralViewer} , which can for reducing patients' fear about the surgery. 
% mainly because the intuitive view of oral cavity in 3D (E1,E2,E3), visualizing instruments (E3) and animating their effects in real-time (E1,E2,E3). \xac{these feedbacks are repetitive to those in demo. effect?}
The experts also rated agreement of 5.00 out of 7 for integrating \textit{OralViewer} into their existing practices.  
As mentioned by E2, the tool can be very necessary with the patients' recently growing need for improved dentist visit experience and their willingness to involve in treatment planning.
He also pointed out that animating procedures on patient-specific model can possibly contribute to higher patient satisfaction because of the personalized communications. 
While agreeing on \textit{OralViewer}'s effectiveness of patient education, E3 mentioned that an improved virtual instrument control design can gain a higher preference from dentists for more fluently utilizing the system. 
% \xac{this reads very similar to the previous two RQs; maybe should remove RQ4? or merge its findings into RQ2 and 3?}

\section{Limitations, Discussions, and Future Work}
\subsection{Improving Oral Cavity Modeling}
We enable 3D oral cavity modeling by assembling patient-specific teeth structures estimated from X-ray and registered gum and jaw bone templates. 
However, two improvements can be made according to interviews from the expert study
\eg the current gum templates as the smooth volumes embodying a set of CT-extracted jaw bones are reported to be coarse in form (E3).
Although such approximation is sufficient for patient education, E3 suggested that improved gum templates should be pre-built from an existing intra-oral scanning, which is a dental imaging modality that is capable of capturing soft tissue. 
Furthermore, the current reconstructed teeth do not model root canals, which can be useful in certain surgeries, \eg root canal treatment (E3). 
Future work can enable the root canal modeling by either augmenting the current solid teeth model with artifact canals or including the root canal modeling in the ConvNet training process. 

\subsection{Extending Virtual Instrument Set for All Dental Surgeries} 
The current implementation of \textit{OralViewer} includes 6 common dental instruments, which can be applied to conduct various virtual surgeries including wisdom tooth extraction, apicoectomy, crown restoration and lengthening, \textit{etc.} 
However, more virtual instruments are required towards a comprehensive system for all dental surgeries, \eg endodontic files for root canal treatment. 
Since most virtual instrument effect can be simulated with CSG (Constructive Solid Geometry) operations, \eg Boolean operation between two models, and mesh sculpting, \eg smoothing, creasing, and flattening, we suggest that \textit{OralViewer} should be extended to allow dentists to register new virtual dental instruments. 
Specifically, various CSG and mesh sculpting functions can be implemented and a dentist can add a new instrument by importing a instrument model for visualization and its parameters for configuring the operating effects.
% , and (\textit{ii}) selecting the animation effect of the instrument from the pool. 

\subsection{Improving Dental Instrument Visualization} 
\textit{OralViewer} visualizes a model of the selected dental instrument for enhancing patients' understanding. 
However, expert E3 suggested that the visualization should be improved from two aspects.
First, when the animation effect of the instrument is changed, the appearance of the instrument can be dynamically altered to reflect the effect change. 
For example, the head of a handpiece can become larger when the effecting size of grinding is set to be larger. 
Second, different models of an instrument can be pre-defined for dentists' selection to reflect the real surgery situation. 
An example as E3 mentioned is that, there are multiple periapical curette tips of different shapes, each of which is applied according to periapical cavity access condition in real surgeries.

\bibliographystyle{ACM-Reference-Format}

\typeout{get arXiv to do 4 passes: Label(s) may have changed. Rerun}
\end{document}